\documentclass{sigchi-ext}
\usepackage[T1]{fontenc}
\usepackage{textcomp}
\usepackage[scaled=.92]{helvet} 
\usepackage{graphicx} 
\usepackage{balance}  
\usepackage{booktabs} 
\usepackage{ccicons}  
\usepackage{ragged2e} 

\usepackage[all]{nowidow}

\usepackage[utf8]{inputenc} 

\def\plaintitle{Remote Biofeedback Sharing, Opportunities and Challenges}
\def\plainauthor{Jeremy Frey, Jessica R. Cauchard}
\def\plainkeywords{Biofeedback, Tangible interface, Remote Communication}

\title{\plaintitle}

\numberofauthors{3}
\author{%
  \alignauthor{%
    \textbf{Jérémy Frey}\\
    \affaddr{IDC Herzliya, Israel}\\
    \affaddr{Ullo, France}\\
    \email{jfrey@ullo.fr}}  \alignauthor{%
    \textbf{Jessica R. Cauchard}\\
    \affaddr{IDC Herzliya, Israel}\\
    \email{jcauchard@acm.org}} \\
 }

\definecolor{linkColor}{RGB}{6,125,233}
\hypersetup{%
  pdftitle={\plaintitle},
  pdfauthor={\plainauthor},
  pdfkeywords={\plainkeywords},
  bookmarksnumbered,
  pdfstartview={FitH},
  colorlinks,
  citecolor=black,
  filecolor=black,
  linkcolor=black,
  urlcolor=linkColor,
  breaklinks=true,
}

\copyrightinfo{Permission to make digital or hard copies of part or all of this work for personal or classroom use is granted without fee provided that copies are not made or distributed for profit or commercial advantage and that copies bear this notice and the full citation on the first page. Copyrights for third-party components of this work must be honored. For all other uses, contact the Owner/Author.\\
Copyright is held by the owner/author(s).\\
{\emph{UbiComp/ISWC'18 Adjunct}}, October 8-12, 2018, Singapore, Singapore.\\
ACM 978-1-4503-5966-5/18/10. \\
https://doi.org/10.1145/3267305.3267701}

\begin{document}

\teaser{
  \centering
  \includegraphics[width=1\textwidth]{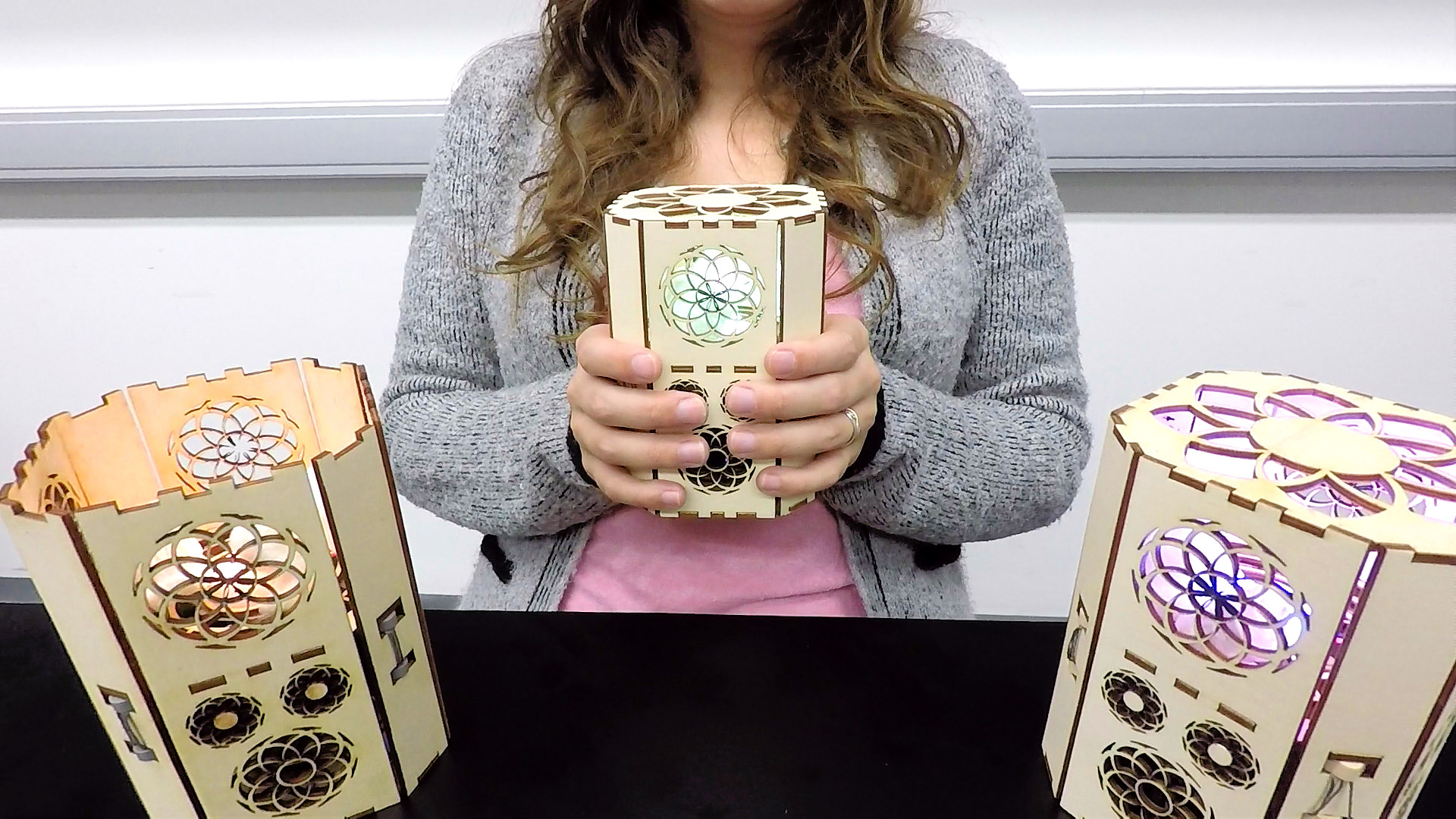}
  \caption{Dišimo: an ambient and shared biofeedback about heart rate variability \protect\cite{Mladenovic2018}.\label{fig:dishimo}}
}

\maketitle

\RaggedRight{} 


\begin{abstract}

Biofeedback is commonly used to regulate one's state, for example to manage stress. The underlying idea is that by perceiving a feedback about their physiological activity, a user can act upon it. In this paper we describe through two recent projects how biofeedback could be leveraged to share one's state at distance. Such extension of biofeedback could answer to the need of belonging, further widening the applications of the technology in terms of well-being.

\end{abstract}



\section{Introduction}

Maslow describes in \cite{Maslow1943} a hierarchy of needs, a model of states and goals that one would strive to reach throughout their life.
While this model was debated and refined over the years, it is still a strong framework to describe a trajectory that could lead to healthiness and well-being. For instance, in the current paper, we use this framework to describe how and why biofeedback could be employed to increase well-being. Moreover, we argue that biofeedback could be used to leverage higher levels of Maslow's hierarchy. 

Biofeedback is a method that enables users to learn to control autonomous bodily processes. It relies on physiological measurements; most of the time, signals originate from respiration, electrodermal activity or heart rate \cite{McKee2008}. These measures are then be translated to an output that the user is able to perceive, for example a visual or an audio cue. While this physiological activity is mostly involuntary and is associated with a low level of awareness, through biofeedback it becomes possible to visualize, reflect and eventually act on it in real-time. 

Biofeedback applications are diverse, from rehabilitation to stress management \cite{McKee2008}. For instance, someone could use a combination of breathing guidance and breathing biofeedback to reach a specific respiration and breathe ``correctly''. That situation could be associated with the first ``level'' in the hierarchy of needs, Physiologycal needs. However this situation is also closely associated to the second level, Safety needs. Indeed, when a stressful stimuli arise, basic physiological responses will occur -- i.e. ``fight or flight'' responses -- and, in case of a pathological state of anxiety, the sole \emph{anticipation} of those stimuli is sufficient to trigger stress, without their actual appearance. Therefore, by being aware at the earliest of those changes, one could regulate themselves their state and alleviate the suffering. 

In the following sections, we demonstrate ways to put biofeedback into practice through two recent projects. Dišimo and Breeze are two tangible devices drawing from ubiquitous computing in order to propose an ``ecological'' biofeedback that could be integrated in the environment of the users. Specifically, they were designed to enable different levels of connectedness. By sharing signals among devices, it becomes possible to supplement and positively influence the third need in the Maslow pyramid: Belonging. 

\section{Dišimo}

Dišimo \cite{Mladenovic2018} acts as a gentle and ambient feedback of one's state (Figure \ref{fig:dishimo}). It provides a multi-modal feedback through light, sounds, and the actuation of physical particles. When used in conjunction with a smartwatch measuring heart rate, it will sense a decrease in heart rate variability (HRV) -- a sign of stress or cognitive workload \cite{Fairclough2004} -- and play a sound as a reminder to breathe. If the user chooses to ignore the device, it will automatically stop playing the sound. Otherwise, if they decide to take a break and grasp Dišimo, the device lights up. Then, upon increase in HRV, physical particles embedded inside Dišimo will start to flutter and produce harmonious tones when hitting the enclosure. The device itself can be equipped with sensors to monitor HRV by the mean of electrocardiography when users grasp its edges, where conductive thread is positioned.

Not only one Dišimo can be used for self-regulation over the course of the day with breathing exercises, but several of them can be connected remotely among a cluster of users. This way, a relaxation session can be experienced between close ones. In this scenario, the global brightness of the light is mapped to the ratio of active users who increased their HRV. In order to be as non-judgmental as possible, we purposely avoided to give information about who specifically reached a higher HRV. Dišimo is not meant to foster competition, that would defeat the purpose of improving well-being, instead it is an aid and a mediator. Knowing that relatives or friends are using Dišimo could be an incentive to use the device, through emulation. Moreover, joining a relaxation session could create an alternate way to connect and empathize with the others, even without physical contact or explicit communication. While this specific project enables a group to occasionally communicate their relaxation, the next project enables a direct and real-time sharing of biofeedback throughout the day.

\section{Breeze}

\begin{figure}
\centering
  \includegraphics[width=1\columnwidth]{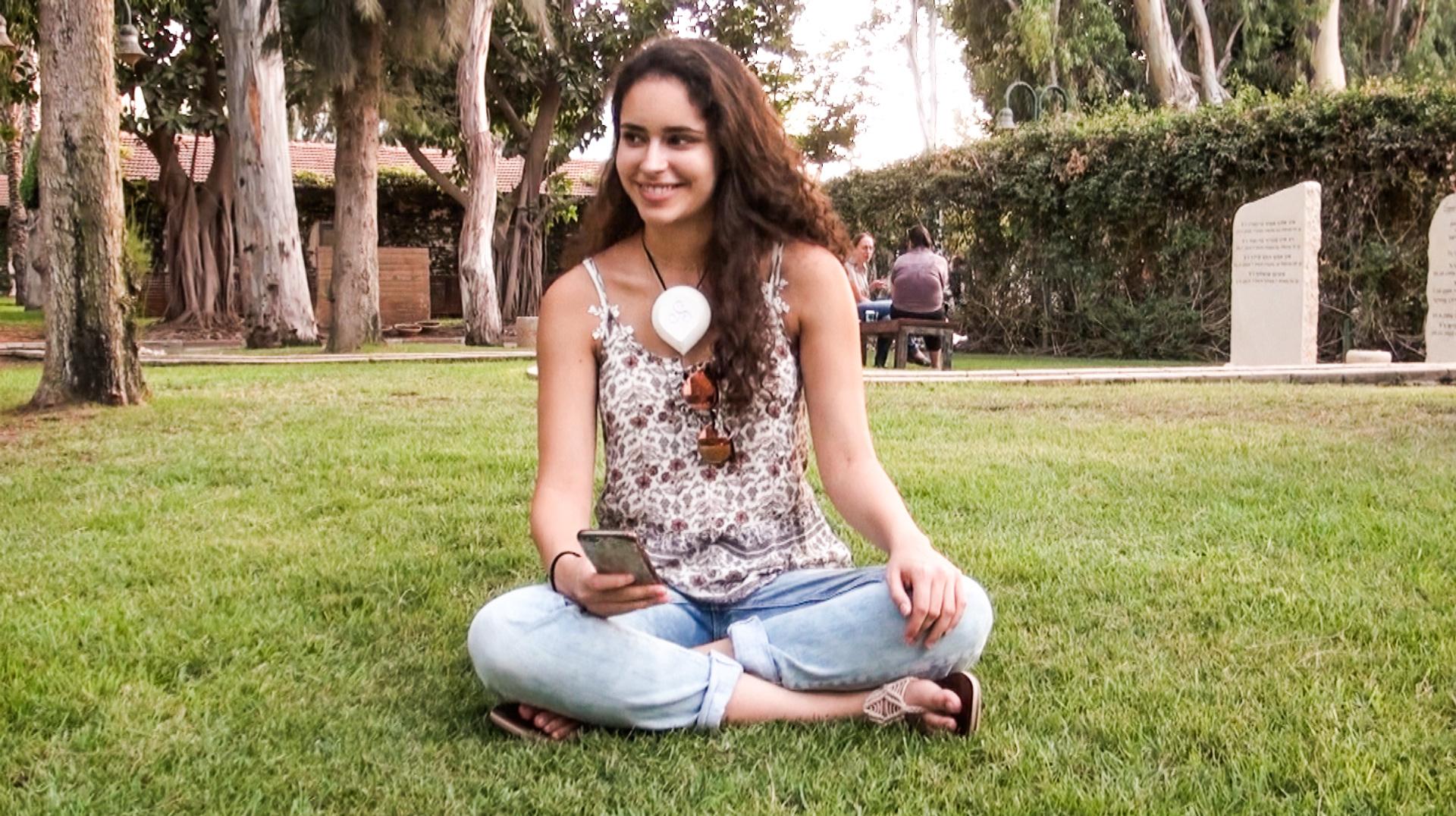}
  \caption{Breeze, a pendant for shared breathing biofeedback \protect\cite{Frey2018}.}\label{fig:breeze}
\end{figure}

Breeze \cite{Frey2018} was developed as a wearable device to communicate breathing biofeedback (Figure \ref{fig:breeze}). Breeze functions bidirectionally, by collecting a person's breathing and providing in real time an ambient biofeedback to a paired pendant. Three biofeedback modalities are embedded in the device: visual, audio and haptic. Each one maps the entire range of breathing to one degree of freedom (i.e. light brightness, pink noise amplitude, vibration motor intensity), thus supporting a more natural and intuitive interface. The wearable relies on an inertial measurement unit to measure breathing, a technology both portable and non-invasive. Breeze is worn as a pendant; this location reduces motion artifacts and increases comfort, while each feedback modality can still be perceived.

Compared to other physiological signals, breathing can be easily modulated. It also conveys a variety of information; for instance many features could be associated to different emotions. During a lab study, we demonstrated how users wearing Breeze were able to perceive and discriminate dimensions of emotions from various breathing patterns, even when they were conveyed through a rather simple biofeedback.
Interestingly, almost all participants were mimicking the breathing pattern in order to understand it. We were surprised by how empathetic participants became toward unimodal cues. This is an encouraging finding; it might indicate that when exposed to such a shared biofeedback, users have a tendency to naturally -- and physically -- engage with it. That could help to create a ``meta biofeedback loop'' between users, with emerging properties to be expected from such interaction.

Results gathered from the study reinforce the hypothesis that by sharing biofeedback it could be possible to increase the feeling of belonging. In fact, participants described how they would use Breeze in long distance relationships with people close to them, such as a partner or family member. It should be noted that in general participants would only use such device with loved ones, probably since feeling someone else's breathing pertains to strong intimacy. We envision that making physiological signals more visible could promote empathy, enforcing a bond between people. However, no matter how grand, this opportunity to extend on existing biofeedback applications comes with its challenges, among which are privacy and acceptability.


\section{Challenges}

Physiological sensors are constantly being refined, up to the point that it is now possible to remotely measure physiological states without the awareness of the person being monitored. For example one could use wireless signals in the Wifi range to detect breathing, \emph{even when the person is behind a wall} (e.g. \cite{Ravichandran2015}). Besides its convenience, such technology questions users' consent and privacy. One of the core design principle of our projects is that users have ownership of their physiological activity. It is up to them to measure and stream their signals to others. On the other hand, when users perceive the biofeedback of someone else, for example with Breeze, it is because their partner chose to activate the device.
We stumbled upon another aspect of privacy while we were developing our prototypes. In a scenario where one would be willing to share a biofeedback with a relative throughout the day, the question of the feedback modality arises. Depending on the social context -- family, friends, co-workers, strangers -- users might not want outsiders to be aware of this particular bond. This is where ownership and customization are important. In Breeze there are three different output modalities that users can freely choose from (visual, audio, haptic). Each possesses a specific ``radius'' and can be perceived to a different extend by self and others depending on the choice of the user, so as not to impede social acceptance.

Within the two projects hereby presented, users have a symmetric and bidirectional relationship; all devices are equals and they present incentives to share with our peers. This use case brings new challenges. In particular, some expectations might arise when people are ``connected'' like so; one sending a biofeedback might expect to receive a feedback in return, or one might be tempted to listen to a biofeedback without sharing. These (mis)use of the technology can be alleviated, though; as for example with Dišimo where by design it is not possible to know exactly who in a cluster participates to a relaxation session. We will argue that freedom and reciprocity are important to empower people and ensure a balanced interaction, and that these considerations should be embedded in the design of a shared biofeedback so as not to be detrimental to well-being.

\section{Conclusion}

Over the course of this paper we presented two examples of shared biofeedback that were designed not only to increase awareness and help users to regulate their inner states, but that can as well foster a feeling of connectedness. Using Maslow's hierarchy of needs, it is straightforward to articulate the underlying mechanisms and understand how such devices can positively impact well-being. While there are challenges associated to the technology -- both in terms of practical aspects and users' empowerment -- a careful design can mitigate the appearance of unforeseen negative effects. Our next step is to deploy this technology outside the laboratory during longitudinal studies, in order to asses how such artifacts will be employed in the long run, and how they will affect people's life and interactions.


\balance{}
\bibliographystyle{SIGCHI-Reference-Format}
\bibliography{biblio}

\end{document}